# On finding minimal $w$-cutset

**Bozhena Bidyuk** and **Rina Dechter**
Information and Computer Science
University Of California Irvine
Irvine, CA 92697-3425

## Abstract

The complexity of a reasoning task over a graphical model is tied to the induced width of the underlying graph. It is well-known that the conditioning (assigning values) on a subset of variables yields a subproblem of the reduced complexity where instantiated variables are removed. If the assigned variables constitute a cycle-cutset, the rest of the network is singly-connected and therefore can be solved by linear propagation algorithms. A $w$-cutset is a generalization of a cycle-cutset defined as a subset of nodes such that the subgraph with cutset nodes removed has induced-width of $w$ or less. In this paper we address the problem of finding a minimal $w$-cutset in a graph. We relate the problem to that of finding the minimal $w$-cutset of a tree-decomposition. The latter can be mapped to the well-known *set multi-cover* problem. This relationship yields a proof of NP-completeness on one hand and a greedy algorithm for finding a $w$-cutset of a tree decomposition on the other. Empirical evaluation of the algorithms is presented.

## 1 Introduction

A cycle-cutset of an undirected graph is a subset of nodes that, when removed, the graph is cycle-free. Thus, if the assigned variables constitute a cycle-cutset, the rest of the network is singly-connected and can be solved by linear propagation algorithms. This principle is at heart of the well-known cycle-cutset conditioning algorithms for Bayesian networks [16] and for constraint networks [7]. Recently, the idea of cutset-conditioning was extended to accommodate search on any subset of variables using the notion of $w$-cutset, yielding a hybrid algorithmic scheme of conditioning and inference paramterized by $w$ [18]. The $w$-cutset is defined as a subset of nodes in the graph that, once removed, the graph has tree-width of $w$ or less.

The hybrid *w-cutset-conditioning* algorithm applies search to the cutset variables and exact inference (e.g., bucket elimination [8]) to the remaining network. *Given a $w$-cutset $C_w$, the algorithm is space exponential in $w$ and time exponential in $w + |C_w|$* [9]. The scheme was applied successfully in the context of satisfiability [18] and constraint optimization [13]. More recently, the notion of conditioning was explored for speeding up sampling algorithms in Bayesian networks in a scheme called *cutset-sampling*. The idea is to restrict sampling to $w$-cutset variables only (peform inference on the rest) and thus reduce the sampling variance ([5, 6]).

Since the processing time of both search-based and sampling-based schemes grows with the size of the $w$-cutset, it calls for a secondary optimization task for finding a minimal-size $w$-cutset. Also, of interest is the task of finding the full sequence of minimal $w$-cutsets, where $w$ ranges from 1 to the problem's induced-width (or tree-width), so that the user can select the $w$ that fits his/her resources. We call the former the $w$-*cutset problem* and the latter *the sequence $w$-cutset problem*. The $w$-cutset problem extends the task of finding minimum cycle-cutset (e.g. a 1-cutset), a problem that received fair amount of attention [3, 2, 20].

The paper addresses the minimum size $w$-cutset and, more generally, the minimum weight $w$-cutset problem. First, we relate the size of a $w$-cutset of a graph to its tree-width and the properties of its tree-decompositions. Then, we prove that the problem of finding a minimal $w$-cutset of a given tree decomposition is NP-complete by reduction from the *set multi-cover* problem [20]. Consequently, we apply a well-known greedy algorithm (GWC) for the set multi-cover problem to solve the minimum $w$-cutset problem. The algorithm finds $w$-cutset within $O(1+\ln m)$ of optimal where $m$ is the maximum number of clusters of size greater than $w + 1$ sharing the same variable in the input tree decomposition. We investigate its performance empirically and show that, with rare exceptions, GWC and its variants find a smaller $w$-cutset than the well-performing MGA cycle-cutset algorithm [3] (adapted to the $w$-cutset problem) and a $w$-cutset algorithm (DGR) proposed in [11].

## 2 Preliminaries and Background

### 2.1 General graph concepts

We focus on automated reasoning problems $R=<X,F>$ where $X$ is a set of variables and $F$ is a set of functions over subsets of the variables. Such problems are associated with a graphical description and thus also called *graphical models*. The primary examples are constraint networks and Bayesian or belief networks formally defined in [16]. The structure of a reasoning problem can be depicted via several graph representations.

DEFINITION **2.1 (Primal-, dual-,hyper-graph of a probem)** *The* primal *graph* $G=<X,E>$ *of a reasoning problem $<X,F>$ has the variables $X$ as its nodes and an arc connects two nodes if they appear in the scope of the same function $f \in F$. A* dual *graph of a reasoning problem has the scopes of the functions as the nodes and an arc connects two nodes if the corresponding scopes share a variable. The arcs are labelled by the shared variables. The* hypergraph *of a reasoning problem has the variables $X$ as nodes and the scopes as edges. There is a one-to-one correspondance between the hypergraph and the dual graphs of a problem.*

We next describe the notion of a tree-decomposition [10, 14]. It is applicable to any grapical model since it is defined relative to its hypergraph or dual graph.

DEFINITION **2.2 (tree-decomp., cluster-tree, tree-width)** *Let $R = <X, D, F>$ be a reasoning problem with its hypergraph $\mathcal{H} = (X, F)$). (We abuse notation when we identify a function with its scope). A tree-decomposition for R (resp., its hypergraph $\mathcal{H}$) is a triple $<T, \chi, \psi>$, where $T=<V, E>$ is a tree, and $\chi$ and $\psi$ are labeling functions which associate with each vertex $v \in V$ two sets, $\chi(v) \subseteq X$ and $\psi(v) \subseteq F$ such that*

1. *For each function $f_i \in F$, there is* exactly *one vertex $v \in V$ such that $f_i \in \psi(v)$, and $scope(f_i) \subseteq \chi(v)$.*

2. *For each variable $X_i \in X$, the set $\{v \in V | X_i \in \chi(v)\}$ induces a connected subtree of $T$. This is also called the running intersection property.*

We will often refer to a node and its functions as a *cluster* and use the term *tree-decomposition* and *cluster tree* interchangeably. The maximum size of node $\chi_i$ minus 1 is the width of the tree decomposition. The *tree width* of a graph $G$ denoted $tw(G)$ is the minimum width over all possible tree decompositions of G [1].The tree-width of a graph is also referred to as *induced width* of a graph. We sometime denote the optimal tree-width of a graph by $tw*$.

There is a known relationship between tree-decompositions and chordal graphs. A graph is *chordal* if any cycle of length 4 or more has a chord. Every tree-decomposition of a graph corresponds to a chordal graph that augments the input graph where the tree clusters are the cliques in the chordal graph. And vice-versa, any chordal graph is a tree-decomposition where the clusters are the maximal cliques. Therefore, much of the discussion that follows, relating to tree-decompositions of graphs, can be rephrased using the notion of chordal graph embeddings.

### 2.2 $w$-cutset of a graph

DEFINITION **2.3 ($w$-cutset of a graph)** *Given a graph $G=<X,E>$, $C_w \subset X$ is a $w$-cutset of G if the subgraph over $X \backslash C$ has tree-width $\leq w$. Clearly, a $w$-cutset is also a $w'$-cutset when $w' \geq w$. The cutset $C_w$ is* minimal *if no $w$-cutset of smaller size exists.*

For completeness, we also define the weighted $w$-cutset problem that generalizes minimum $w$-cutset problem (where all node weights are assumed the same). For example, in $w$-cutset conditioning, the space requirements of exact inference are $O(d_{max}^w)$ where $d_{max}$ is the maximum node domain size in graph $G$. The total time required to condition on $w$-cutset $C$ is $O(N \cdot d_{max}^w) \times |D(C)|$ where $|D(C)|$ is the size of the cutset domain space and N is the number of variables. The upper bound value $d_{max}^{|C|}$ on $|D(C)|$ produces a bound on the computation time of the cutset-conditioning algorithm: $O(N \cdot d_{max}^w) \times d_{max}^{|C|} = O(N \cdot d_{max}^{w+|C|})$. In this case, clearly, we want to minimize the size of C. However, a more refined optimization task is to minimize the actual value of $|D(C)|$:

$$|D(C)| = \prod_{C_i \in C} |D(C_i)|$$

Since the minimum of $|D(C)|$ corresponds to the minimum of $\lg |D(C)|$, we can solve this optimization task by assigning each node $X_i$ cost $c_i = \lg |D(X_i)|$ and minimizing the cost of cutset:

$$cost(C) = \lg |D(C)| = \sum_{C_i \in C} \lg |D(X_i)| = \sum_i c_i$$

Similar considerations apply in case of the $w$-cutset sampling algorithm. Here, the space requirements for the exact inference are the same. The time required to sample a node $C_i \in C$ is $O(d_{max}^w) \times |D(C_i)|$. The total sampling time is $O(d_{max}^w) \times \sum_{C_i \in C} |D(C_i)|$. To minimize the total processing time, we assign each node $X_i$ cost $c_i = |D(X_i)|$ and select the $w$-cutset of minimum cost:

$$cost(C) = \sum_{C_i \in C} |D(C_i)|$$

DEFINITION **2.4 (weighted $w$-cutset of a graph)** *Given a reasoning problem $<X, F>$ where each node $X_i \in X$ has associated $cost(X_i) = c_i$, the cost of a $w$-cutset $C_w$ is given by: $cost(C_w) = \sum_{X_i \in C_w} c_i$. The minimum weight $w$-cutset problem is to find a min-cost $w$-cutset.*

In practice, we can often assume that all nodes have the same cost and solve the easier minimal $w$-cutset problem. In section 3, we establish relations between the size of $w$-cutset of a graph and the width of its tree-decomposition. In section 4, we show that the problem is NP-hard even when finding a minimum $w$-cutset of a chordal graph (corresponding to a tree-decomposition of a graph).

## 3 $w$-Cutset and Tree-Decompositions

In this section, we explore relationship between $w$-cutset of a graph and its tree-decomposition.

**THEOREM 3.1** *Given a graph $G=<X,E>$, if $G$ has a $w$-cutset $C_w$, then there is a tree-decomposition of $G$ having a tree-width* tw $\leq |C_w| + w$.

**Proof.** If there exists a $w$-cutset $C_w$, then we can remove $C_w$ from the graph yielding, by definition, a subgraph $G'$ over $X \backslash C_w$ that has a tree decomposition T with clusters of size at most w+1. We can add the set $C_w$ to each cluster of T yielding a tree-decomposition with clusters of size at most $w + 1 + |C_w|$ and tree-width $w + |C_w|$. □

We can conclude therefore that for any graph $tw^* \leq |C_i| + i$ for every $i$. Moreover,

**THEOREM 3.2** *Given a graph $G$, if $c_i^*$ is the size of a smallest $i$-cutset $C_i^*$, and $tw^*$ is its tree-width, then:*
$$c_1^* + 1 \geq c_2^* + 2 \geq ... \geq c_i^* + i \geq ... \geq tw^* \quad (1)$$

**Proof.** Let us define $\Delta_{i,i+1} = c_i^* - c_{i+1}^*$, then we claim that $\Delta_{i,i+1} \geq 1$. Assume to the contrary that $c_i = c_{i+1}$, that is $\Delta_{i,i+1}$=0. Since $C_i^*$ is an i-cutset, we can build a tree decomposition T with maximum cluster size (i+1). Note, for all $i < tw^*$, $|C_i^*| > 0$ or we would have $tw^* = i$. Pick some $X_j \in C_i^*$ and add $X_j$ to every cluster yielding tree decomposition $T'$ with maximum cluster size (i+2). Clearly, $C_i^* \backslash X_j$ is an (i+1)-cutset of size $c_i^* - 1 = c_{i+1}^* - 1$ which contradicts the minimality of $C_{i+1}^*$. □

Given a graph $G = (V, E)$, the $w$-*cutset sequence* problem seeks a sequence of minimal $j$-cutsets where $j$ ranges from 1 to the graph's tree-width: $C_1^*, ..., C_j^*, ..., C_{tw^*} = \phi$. Let $C'_w$ be a subset-minimal $w$-cutset, namely one that does not contain another $w$-cutset. If we have a $w$-cutset sequence, we can reason about which $w$ to choose for applying the $w$-cutset conditioning algorithm or $w$-cutset sampling. Given a $w$-cutset sequence we define a function $f(i) = |C_i| + i$ where $i$ ranges from 1 to $tw$. This function characterizes the complexity of the $w$-cutset conditioning algorithms: for each $i$, the space complexity is exponential in $i$ and the time complexity is exponential in $f(i)$. The space consideration suggests selecting $i$ as small as possible. Notice that for various intervals of $i$, $f(i)$ maybe constant: if $|C_i| = |C_{i+1}| + 1$. Thus, given a $w$-cutset sequence, whenever $f(i) = f(i + 1)$, then $w = i$ is preferred over

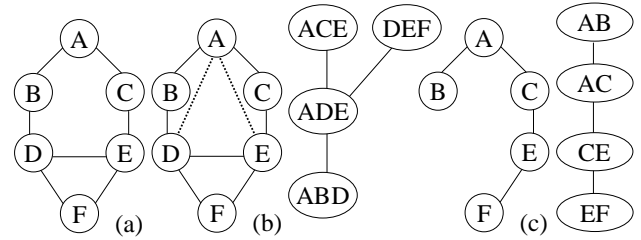

Figure 1: (a) Graph; (b) triangulated graph and corresponding tree decomposition of width 2; (c) graph with 1-cutset node $\{D\}$ removed and corresponding tree-decomposition.

$w = i + 1$. Alternatively, given a bound on the space-complexity expressed by $r$, we can select a most preferred $w_p$-cutset such that:
$$w_p(r) = \arg\min_j \{r = j\}$$

In the empirical section 6, we demonstrate the analysis of function $f(i)$ and its implications.

**THEOREM 3.3** *Given a tree-decomposition $T=(V,E)$ where $V=\{V_1, ..., V_t\}$ is the set of clusters and given a constant $w$, a minimum $w$-cutset $C_w^*$ of $G$ satisfies:*
$$|C_w^*| \leq \sum_{i,|V_i|>w+1} (|V_i| - (w+1)) \quad (2)$$

**Proof.** From each cluster $V_i \in V$ of size larger than (w+1), select a subset of nodes $C_i \subset V_i$ of size $|C_i| = |V_i| - (w + 1)$ so that $|V_i \backslash C_i| \leq w + 1$. Let $C_w = \cup_{i,|V_i|>w+1} C_i$.
By construction, $C_w$ is a $w$-cutset of $G$ and:
$c_w^* \leq |C_w| = |\cup_i C_i| \leq \sum_i |C_i| = \sum_{i,|V_i|>w+1} |V_i| - (w+1)$. □

Since a $w$-cutset yields a tree-decomposition having $tw = w$, it looks reasonable when seeking $w$-cutset to start from a good tree-decomposition and find its $w$-cutset (or a sequence). In particular, this avoids the need to test if a graph has $tw = w$. This task is equivalent to finding a $w$-cutset of a chordal (triangulated) graph.

**DEFINITION 3.1 (A $w$-cutset of a tree-decomposition)**
*Given a tree decomposition $T=<V,E>$ of a reasoning problem $<X,F>$ where $V$ is a set of subsets of $X$ then $C_w^T \subset X$ is a $w$-cutset relative to $T$ if for every $i$, $|V_i \backslash C_w^T| \leq w + 1$.*
We should note upfront, however, that a minimum-size $w$-cutset of $T$ (even if $T$ is optimal) is not necessarily a minimum $w$-cutset of $G$.

**Example 3.4** *Consider a graph in Figure 1(a). An optimal tree decomposition of width 2 is shown in Figure 1(b). This tree-decomposition clearly does not have a 1-cutset of size $< 2$. However, the graph has a 1-cutset of size 1, $\{D\}$, as shown in Figure 1(c).*

On the other hand, given a minimum $w$-cutset, removing the $w$-cutset from the graph yields a graph having $tw^* = w$. Because, otherwise, there exists a tree-decomposition over $X \backslash C_w$ having $tw < w$. Select such a tree and select a node in $C_w$ that can be added to the tree-decomposition without increasing its tree-width beyond $w$. Such a node must exist, contradicting the minimality of $C_w$.

It is still an open question if every minimal $w$-cutset is a $w$-cutset of some minimum-width tree-decomposition of $G$.

## 4 Hardness of $w$-Cutset on Cluster-Tree

While it is obvious that the general $w$-cutset problem is NP-complete (1-cutset is a cycle-cutset known to be NP-complete), it is not clear that the same holds relative to a given tree-decomposition. We now show that, given a tree-decomposition T of a hyper-graph $\mathcal{H}$, the $w$-cutset problem for $T$ is NP-complete. We use a reduction from *set multi-cover* (SMC) problem.

**DEFINITION 4.1 (Set Cover (SC))** *Given a pair $<U,S>$ where $U$ is universal set and $S$ is a set of subsets $S=\{S_1,...,S_m\}$ of $U$, find a minimum set $C \subset S$ s.t. each element of $U$ is covered at least once: $\cup_{S_i \in C} S_i = U$.*

**DEFINITION 4.2 (Set Multi-Cover(SMC))** *Given a pair $<U,S>$ where $U$ is universal set and $S$ is a set of subsets $S = \{S_1,...,S_m\}$ of $U$, find a minimum cost set $C \subset S$ s.t. each $U_i \in U$ is covered at least $r_i > 0$ times by C.*

The SC is an instance of SMC problem when $\forall i, r_i = 1$.

**THEOREM 4.1 (NP-completeness)** *The problem "Given a tree-decomposition $T=<V,E>$ and a constant $k$, does there exist a $w$-cutset of $T$ of size at most $k$ ?" is NP-complete.*

**Proof.** Given a tree decomposition $T=<V,E>$ over X and a subset of nodes $C \in X$, we can verify in linear time whether C is a $w$-cutset of T by checking if $\forall V_i \in V$, $|V_i \backslash C| \leq w+1$. Now, we show that the problem is NP-hard by reduction from set multi-cover.

Assume we are given a set multi-cover problem $<U,S>$, where $U=\{X_1,...,X_n\}$ and $S=\{S_1,...,S_m\}$, a covering requirement $r_i > 0$ for each $U_i \in U$.

We define a cluster tree $T=<V,E>$ over $S$ where there is a node $V_i \in V$ corresponding to each variable $U_i$ in U that contains all subsets $S_j \in S$ that cover node $X_i$: $V_i = \{S_j \in S|X_i \in S_j\}$. Additionally, there is a node $V_S \in V$ that contains all subsets in S: $V_S = S$. Thus, $V = \{V_i|U_i \in U\} \cup V_S$. The edges are added between each cluster $V_{i,i \neq s}$ and cluster $V_S$: $E = \{V_i V_S | U_i \in U\}$ to satisfy running intersection property in $T$.

Define $w+1=|S| - \min_i r_i = m - \min_i r_i$. For each $V_{i,i \neq s}$,

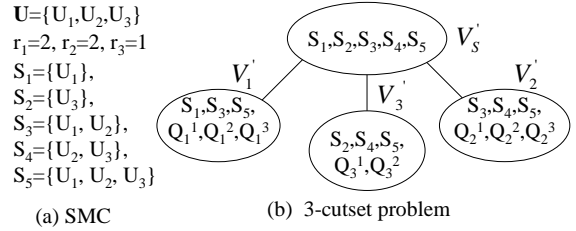

(a) SMC  (b) 3-cutset problem

Figure 2: (a) A set multi-cover problem $<U,S>$ where $U=\{U_1,U_2,U_3\}$, $S=\{S_1,...,S_5\}$, the covering requirements $r_1$=2, $r_2$=2, $r_3$=1. (b) Corresponding augmented tree decomposition $T'=<V',E>$ over $S'=\{S_1,...,S_5,Q_1^1,Q_1^2,Q_1^3,Q_2^1,Q_2^2,Q_2^3,Q_3^1,Q_3^2\}$.

since $|V_i| \leq m$ and $r_i > 0$, then $|V_i|-r_i \leq |V_i|-\min_i r_i \leq m - \min_i r_i = w+1$. Consequently, $|V_i| \leq r_i + w + 1$.

For each $V_i$ s.t. $|V_i| < r_i + w + 1$, define $\Delta_i = r_i + w + 1 - |V_i|$ and augment cluster $V_i$ with a set of nodes $\mathcal{Q}_i = \{Q_i^1...Q_i^{\Delta_i}\}$ yielding a cluster $V_i' = V_i \cup \mathcal{Q}_i$ of size $|V_i'|=r_i+w+1$.

We will show now that a set multi-cover $<U,S>$ has a solution of size $k$ iff there exists a $w$-cutset of augmented tree decomposition $T'=<V',E>$ of the same size. The augmented tree for sample SMC problem in Figure 2(a) is shown in Figure 2(b).

Let $C$ be a set multi-cover solution of size $k$. Then, $\forall U_i \in U, |C \cap V_i'| \geq r_i$ which yields $|V_i' \backslash C| \leq |V_i'|-r_i = w+1$. Since $|C| \geq \min_i r_i$, then $|V_S \backslash C| \leq |V_S| - \min_i r_i = m - \min_i r_i = w+1$. Therefore, $C$ is a $w$-cutset of size $k$.

Let $C_w$ be a $w$-cutset problem of size $k$. If $C_w$ contains a node $Q_j \in \mathcal{Q}_i$, we can replace it with some node $S_p \in V_i'$ without increasing the size of the cutset. Thus, without loss of generality, we can assume $C_w \subset S$. For each $V_i'$ corresponding to some $U_i \in U$, let $C_i = C_w \cap V_i'$. By definition of $w$-cutset, $|V_i' \backslash C_w| \leq w+1$. Therefore, $|C_i| \geq |V_i'| - (w+1) = r_i$. By definition, $C_w$ is a cover for the given SMC problem.

Minimum $w$-cutset problem is NP-hard by reduction from set multi-cover and is verifiable in linear time. Therefore, minimum $w$-cutset problem is NP-complete. □

**Example 4.2** *Let us demonstrate those steps for the SMC problem with $U=\{U_1,U_2,U_3\}$ and $S=\{S_1,...,S_5\}$ shown in Figure 2(a). Define $T=<V,E>$, $V=\{V_1,V_2,V_3,V_s\}$, over S:*
$V_1=\{S_1,S_2,S_3\}$, $f_1$=3, $V_2=\{S_3,S_4,S_5\}$, $f_2$=3,
$V_3=\{S_2,S_4,S_5\}$, $f_3$=3, $V_S=\{S_1,...,S_5\}$, $f_S$=5.
*Then, $w = |S| - 1 - \min_i r_i = 5 - 1 - 1 = 3$. Augment:*
$V_1$: $\Delta_1 = w+1+r_1 - f_1 = 4+2-3 = 3$, $Q_1=\{Q_1^1,Q_1^2,Q_1^3\}$.
$V_2$: $\Delta_2 = w+1+r_2 - f_2 = 4+2-3 = 3$, $Q_2=\{Q_2^1,Q_2^2,Q_2^3\}$.
$V_3$: $\Delta_3 = w+1+r_3 - f_3 = 4+1-3 = 2$, $Q_3=\{Q_3^1,Q_3^2\}$.
*The augmented tree decomposition $T'$ is shown in Fig-*

ure 2(b). Any SMC solution such as $C=\{S_3, S_5\}$ is a 3-cutset of T and vice versa.

In summary, we showed that when $w$ is not a constant the $w$-cutset problem is NP-complete. This implies that the $w$-cutset sequence problem over tree-decompositions is hard.

## 5 Algorithm GWC for minimum cost $w$-cutset

Next, we show that the problem of finding $w$-cutset can be mapped to that of finding set multi-cover. The mapping suggests an application of greedy approximation algorithm for set multi-cover problem to find $w$-cutset of a tree decomposition. When applied to a tree decomposition $T=<V,E>$ over $X$, it is guaranteed to find a solution within factor $O(1 + \ln m)$ of optimal where $m$ is the maximum # of clusters of size $> (w + 1)$ sharing the same node. To avoid loss of generality, we consider the weighted version of each problem.

The mapping is as follows. Given any $w$-cutset problem of a tree-decomposition $T=<V,E>$ over X, each cluster node $V_i \in V$ of the tree becomes a node of universal set U. A covering set $S_{X_j}=\{V_i \in V | X_j \in V_i\}$ is created for each node $X_j \in X$. The cost of $S_{X_j}$ equals the cost of $X_j$. The cover requirement is $r_i = |V_i| - (w + 1)$. Covering a node in SMC with a set $S_{X_j}$ corresponds to removing node $X_j$ from each cluster in T. Then, the solution to a set multi-cover is a $w$-cutset of T. Let C be a solution to the SMC problem. For each $U_i \in U$, the set C contains at least $r_i$ subsets $S_{X_j}$ that contain $U_i$. Consequently, since $U_i = V_i$, then $|V_i \cap C| \geq r_i$ and $|V_i \backslash C| \leq |V_i| - r_i = |V_i| - |V_i| + (w + 1) = w + 1$. By definition, C is a $w$-cutset. An example is shown in Figure 3. This duality is important because the properties of SC and SMC problems are well studied and any algorithms previously developed for SMC can be applied to solve $w$-cutset problem.

A well-known polynomial time greedy algorithm exists for weighted SMC [20] that chooses repeatedly set $S_i$ that covers the most "live" (covered less than $r_i$ times) nodes $f_i$ at the cost $c_i$: a set that minimizes the ratio $c_i/f_i$. In the context of $w$-cutset, $f_i$ is the number of clusters whose size still exceeds $(w + 1)$ and $c_i$ is the cost of node $X_i$. As discussed earlier, $c_i$ maybe defined as the size of the domain of node $X_i$ or its log. When applied to solve the $w$-cutset problem, we will refer to the algorithm as GWC (Greedy $W$-Cutset). It is formally defined in Figure 4. We define here the approximation algorithm metrics:

DEFINITION 5.1 **(factor $\delta$ approximation)** *An algorithm $\mathcal{A}$ is a factor $\delta$, $\delta > 0$, approximation algorithm for minimization problem $\mathcal{P}$ if $\mathcal{A}$ is polynomial and for every instance $I \in D_\mathcal{P}$ it produces a solution s such that: $cost(s) \leq \delta * cost_{OPT}(s), \delta > 1$.*

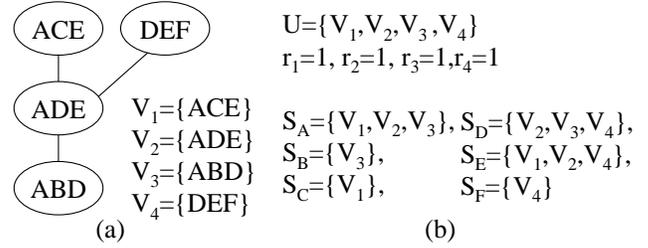

Figure 3: (a) A tree decomposition $T=<V,E>$ where $V=\{V_1,...,V_4\}$ over $X=\{A,B,C,D,E,F\}$; (b) the corresponding set multi-cover problem $<U,S>$ where $U=\{V_1,V_2,V_3,V_4\}$ and $S=\{S_A, S_B, S_C, S_D, S_F\}$; here, set $S_{X_i}$ contains a cluster $V_j$ iff $X_i \in V_j$. The 1-cutset of T is a solution to the set multicover with covering requirements $r_1=r_2=r_3=r_4=1$: when node $V_i \in V$ is "covered" by set $S_{X_i}$, node $X_i$ is removed from each cluster.

---

**Greedy $w$-Cutset Algorithm (GWC)**
**Input:** A set of clusters $V = \{V_1, ..., V_m\}$ of a tree-decomposition over $X = \{X_1, ..., X_n\}$ where $\forall V_i \in V$, $V_i \subset X$; the cost of each node $X_i$ is $c_i$.
**Output :** A set $C \subset X$ s.t. $|V_i \backslash C| \leq w$.
Set $C = \emptyset$, t=0.
**While** $\exists V_i$ s.t. $|V_i| > w + 1$ **do**
1. $\forall X_i \in X$, compute $f_i = |\{V_j\}|$ s.t. $|V_j| > w$ and $X_i \in V_j$.
2. Find node $X_i \in X$ that minimizes the ratio $c_i/f_i$.
3. Remove $X_i$ from all clusters: $\forall V_i \in V, V_i = V_i \backslash X_i$.
4. Set $X = X \backslash X_i, C = C \cup \{X_i\}$.
**End While**
**Return** C

---

Figure 4: Greedy $w$-cutset Algorithm.

GWC is a factor $O(1+\ln m)$ approximation algorithm [17] where $m$ is the maximum number of clusters sharing the same node (same as the maximum set size in SMC).

This bound is nearly the best possible for a polynomial algorithm due to strong inapproximability results for the set cover problem, the special case of set multi-cover problem. Approximation guarantee better than $O(\ln m)$ is not possible for any polynomial time algorithm unless P=NP [4, 15]. Furthermore, $\exists$ absolute constants $C, \Delta_0$ s.t. $\forall \Delta \geq \Delta_0$ no polynomial-time algorithm can approximate the optimum within a factor of $\ln \Delta - C \ln \ln \Delta$ unless P=NP [19].

## 6 Experiments

We use Bayesian networks as input reasoning problems. In all experiments, we started with a moral graph G of a Bayesian network $\mathcal{B}$ which was used as the input to the minimal $w$-cutset problem. The tree-decomposition of G was obtained using min-fill algorithm [12].

Our benchmarks are two CPCS networks from UAI repository: cpcs360b with N=360 nodes and induced width $w*$=22 and cpcs422b with N=422 nodes and induced width $w*$=27, one instance each. Our other benchmarks are layered random networks where each node is assigned a set of parents selected randomly from a previous layer. One set of random networks consisted of 4 layers of $L = 50$ nodes each, total of $N$=50x4=200 nodes; each node assigned $P = 3$ parents. The second set of random networks consisted of 8 layers of $L = 25$ nodes each, total of $N$=25x8=200 nodes; each node assigned $P = 3$ parents. For random networks, the results are averaged over 100 instances. We compare the performance of two greedy heuristic algorithms- MGA (Modified Greedy Algorithm due to [3]) and DGR (Deterministic Greedy Algorithm due to [11])- to our proposed algorithms: GWC (Greedy $W$-Cutset) and its variants. All nodes domains of size 2.

The MGA algorithm is adapted from a minimum cost cycle-cutset algorithm of [3] that iteratively removes all nodes if degree 1 from the graph and adds to cutset the node that minimizes the cost to degree ratio. The algorithm stops when remaining subgraph is cycle-free. The MGA is a factor 4 approximation algorithm. In [20], a factor 2 approximation algorithm is defined based on layering. However, it can not be easily adapted to finding minimal $w$-cutset for $w > 1$. For MGA, the only modification required to find $w$-cutset is to stop when original graph with cutset nodes removed can be decomposed into a cluster tree of width $w$ or less (using min-fill heuristics). In our implementation, MGA algorithm uses the GWC heuristics to break ties: if two nodes have the same degree, the node found in most of the clusters of size $> w$ is added to the cutset.

The DGR algorithm is the Deterministic Greedy Algorithm for finding an elimination order of the variables that yields a tree-decomposition of bounded width defined in [11]. DGR obtains a $w$-cutset while computing the elimination order of the variables. When eliminating some node $X$ yields a cluster that is too large (size $> w + 1$), the algorithm uses greedy heuristics to pick a cutset node among all the nodes that are not in the ordering yet. Specifically, the deterministic algorithm adds to the cutset a node $X$ that maximizes expression $\sqrt{|N_X|}C_X$, where $N_X$ is a set of neighbours of $X$ that are not eliminated yet and $C_X = \prod_{U_i \in N_X} |D(U_i)|$.

The GWC algorithm was implemented as described earlier picking at each iteration a node found in most clusters of size $> w+1$ with a secondary heuristics (tie breaking) that selects the node contained in most of the clusters. Several variants of GWC with different tie breaking heuristics were tested that were allowed to rebuild a tree decomposition after removing a cutset node:

**GWCA** - breaks ties by selecting the node found in most of the clusters of the tree decomposition;
**GWCM** - breaks ties by selecting the node found in most of the clusters of maximum size;

**GWCD** - breaks ties by selecting the node of highest degree (the degree of the node is computed on the subgraph with all cutset nodes removed and all resulting singly-connected nodes removed). Note that GWC and GWCA only differ in that GWCA rebuilds a cluster-tree after removing a cutset node. Also note that MGA and GWCD have their primary and tie-breaking heuristics switched.

Table 1: $w$-cutset. Networks: I=cpcs360b, II=cpcs422b, III=4-layer random networks, L=50, N=200, P=3; IV =8-layer random networks, L=25, N=200, P=3.

|  | $w$ | 1 | 2 | 3 | 4 | 5 | 6 | 7 | 8 | 9 | 10 |
|---|---|---|---|---|---|---|---|---|---|---|---|
| I | MGA | 30 | 22 | 20 | 18 | 16 | 15 | 14 | 13 | 12 | **10** |
| $w*$=20 | DGR | 36 | 22 | 19 | 18 | 16 | **14** | **13** | **12** | **11** | **10** |
|  | GWC | **27** | **20** | **17** | **16** | **15** | **14** | **13** | **12** | **11** | **10** |
|  | GWCA | **27** | 21 | 18 | **16** | **15** | **14** | **13** | **12** | **11** | **10** |
|  | GWCD | **27** | 21 | 18 | **16** | **15** | **14** | **13** | **12** | **11** | **10** |
| II | MGA | 80 | 70 | 65 | 60 | 54 | 49 | 44 | 41 | 38 | 36 |
| $w*$=22 | DGR | 84 | 70 | 63 | 54 | 49 | 43 | 38 | 32 | 27 | 23 |
|  | GWC | **78** | 66 | 58 | 52 | 46 | 41 | 36 | 31 | 26 | 22 |
|  | GWCA | **78** | **65** | **57** | **51** | **45** | **40** | **35** | **30** | **25** | **21** |
|  | GWCD | **78** | **65** | **57** | **51** | **45** | **40** | **35** | **30** | **25** | **21** |
| III | MGA | 87 | 59 | 54 | 52 | 50 | 48 | 47 | 45 | 44 | 43 |
| $w*$=49 | DGR | 80 | 57 | 52 | 50 | 48 | 46 | 44 | 43 | 42 | 40 |
|  | GWC | 78 | 61 | 53 | 49 | 46 | 44 | 43 | 42 | 41 | 39 |
|  | GWCA | **74** | **56** | 50 | **47** | **44** | **42** | **41** | **39** | **38** | **37** |
|  | GWCD | **74** | **56** | **49** | **47** | **44** | **42** | **41** | **39** | **38** | **37** |
| IV | MGA | 99 | 74 | 69 | 66 | 63 | 61 | 59 | 56 | 54 | 51 |
| $w*$=24 | DGR | 90 | 71 | 65 | 61 | 58 | 55 | 52 | 49 | 47 | 44 |
|  | GWC | 93 | 77 | 68 | 63 | 59 | 55 | 52 | 49 | 46 | 43 |
|  | GWCA | 87 | **70** | **62** | **57** | **54** | **51** | **48** | **45** | **42** | **39** |
|  | GWCD | **86** | **70** | **62** | **57** | **54** | **51** | **48** | **45** | **42** | **39** |
|  | $w$ | 11 | 12 | 13 | 14 | 15 | 16 | 17 | 18 | 19 | 20 |
| I | MGA | **9** | **8** | **7** | **6** | **5** | **4** | **3** | **2** | **1** | **0** |
| $w*$=20 | DGR | **9** | **8** | **7** | **6** | **5** | **4** | **3** | **2** | **1** | **0** |
|  | GWC | **9** | **8** | **7** | **6** | **5** | **4** | **3** | **2** | **1** | **0** |
|  | GWCA | **9** | **8** | **7** | **6** | **5** | **4** | **3** | **2** | **1** | **0** |
|  | GWCD | **9** | **8** | **7** | **6** | **5** | **4** | **3** | **2** | **1** | **0** |
| II | MGA | 33 | 30 | 28 | **9** | **8** | 7 | 6 | 5 | 4 | **2** |
| $w*$=22 | DGR | 21 | 19 | 16 | **9** | **8** | 7 | **5** | **4** | **3** | **2** |
|  | GWC | 19 | 16 | 13 | 10 | **8** | 6 | **5** | **4** | **3** | **2** |
|  | GWCA | **18** | **15** | **12** | **9** | **8** | 6 | **5** | **4** | **3** | **2** |
|  | GWCD | **18** | **15** | **12** | **9** | **8** | 6 | **5** | **4** | **3** | **2** |
| III | MGA | 41 | 40 | 39 | 37 | 36 | 35 | 34 | 33 | 31 | 30 |
| $w*$=49 | DGR | 39 | 38 | 36 | 36 | 34 | 33 | 32 | 31 | 30 | 29 |
|  | GWC | 38 | 37 | 36 | 35 | 34 | 33 | 32 | 31 | 30 | 29 |
|  | GWCA | **36** | 35 | **34** | **33** | **32** | **31** | **30** | **29** | **28** | **27** |
|  | GWCD | **36** | **34** | **34** | **33** | **32** | **31** | **30** | **29** | **28** | **27** |
| IV | MGA | 49 | 47 | 44 | 41 | 39 | 36 | 34 | 31 | 28 | 26 |
| $w*$=24 | DGR | 41 | 38 | 36 | 33 | 31 | 28 | 25 | 23 | 21 | 19 |
|  | GWC | 40 | 37 | 35 | 32 | 29 | 27 | 25 | 23 | 20 | 18 |
|  | GWCA | **37** | **34** | **32** | **30** | **27** | **25** | **23** | **21** | **19** | **17** |
|  | GWCD | **37** | 35 | **32** | **30** | 28 | **25** | 24 | **21** | **19** | **17** |

The results are presented in Table 1. For each benchmark, the table provides the five rows of results corresponding to the five algorithms (labelled in the second column). Columns 3-12 are the $w$-cutset sizes for the $w$-value. The upper half of the table entires provides results for $w$ in range $[1, 10]$; the lower half of the table provides results for $w$ in range $[11, 20]$. The results for cpcs360b and cpcs422b correspond to a single instance of each network. The result for random networks are averaged over 100 instances. The best entries for each $w$ are highlighted.

As Table 1 shows, it pays to rebuild a tree decomposition: with rare exceptions, GWCA finds a cutset as small as GWC or smaller. On average, GWCA, GWCM, and GWCD computed the same-size $w$-cutsets. The results for GWCM are omitted since they do not vary sufficiently from the others.

The performance of MGA algorithm appears to depend on the network structure. In case of cpcs360b, it computes the same size $w$-cutset as GWC variants for $w \geq 10$. However, in the instance of cpcs422b, MGA consistently finds larger cutsets except for $w$=20. On average, as reflected in the results for random networks, MGA finds larger cutset than DGR or any of the GWC-family algorithms. In turn, DGR occasionally finds a smaller cutset compared to GWC, but always a larger cutset compared to GWCA and GWCD.

We measured the GWC algorithm approximation parameter $M$ in all of our benchmarks. In cpcs360b and cpcs422b we have $M = 86$ and $M = 87$ yielding approximation factor of $1 + \ln M \approx 5.4$. In random networks, $M$ varied from 29 to 47 yielding approximation factor $\in [4.3, 4.9]$. Thus, if $C$ is the $w$-cutset obtained by GWC and $C_{opt}$ is the minimum size $w$-cutset, then on average:

$$\frac{|C|}{|C_{opt}|} \leq 5$$

Looking at the results as solutions to the sequence $w$-cutset problems, we can inspect the sequence and suggest good $w$'s by analysing the function $f(i) = |C_i| + i$ as described in section 3. To illustrate this we focus on algorithm GWCA for CPC364, CPCS424 and 4-layer random networks (See Table 2).

For cpcs360b we observe a small range of values for $f(i)$, namely $f(i) \in \{20, 21, 23, 28\}$. In this case the point of choice is $w = 4$ because $f(1) = 28, f(2) = 23, f(3) = 21$ while at $i = 4$ we obtain reduction $f(4) = 20$ which stays constant for $i \geq 4$. Therefore, we can have the same time complexity for $w$-cutset as for exact inference ($w* = 20$) while saving a lot in space, reducing space complexity from exponential in 20 to exponential in 4 only. For $w$-cutset sampling this implies sampling 20 variables (out of 360) and for each variable doing inference exponential in 4.

The results are even more interesting for cpcs422b where we see a fast decline in time complexity with relatively slow decline in space complexity for the range $i = 1, ..., 11$. The decline is more moderate for $i \geq 11$ but is still cost-effective: for $i = 16$ we get the same time performance as $i = 20$ and therefore $i = 16$ represents a more cost-effective point.

Table 2: Function $f(i)$ for i=1...16, GWCA. Networks: I=cpcs360b, II=cpcs422b, III=4-layer random, L=50, N=200, P=3.

| | f(i) | | | | | | | | | |
|---|---|---|---|---|---|---|---|---|---|---|
| $i$ | 1 | 2 | 3 | 4 | 5 | 6 | 7 | 8 | 9 | 10 |
| I | 28 | 23 | 21 | 20 | 20 | 20 | 20 | 20 | 20 | 20 |
| II | 79 | 67 | 60 | 55 | 50 | 46 | 42 | 38 | 34 | 31 |
| III | 75 | 57 | 53 | 51 | 49 | 48 | 48 | 47 | 47 | 47 |
| | f(i) | | | | | | | | | |
| $i$ | 11 | 12 | 13 | 14 | 15 | 16 | 17 | 18 | 19 | 20 |
| I | 20 | 20 | 20 | 20 | 20 | 20 | 20 | 20 | 20 | 20 |
| II | 29 | 27 | 25 | 23 | 23 | 22 | 22 | 22 | 22 | 22 |
| III | 47 | 47 | 47 | 47 | 47 | 47 | 47 | 47 | 47 | 47 |

Finally, for the case of 4-layer random networks, on average the function $f(i)$ decreases for $i = 1...8$ and then remains constant. This suggests that if space complexity allows, the best point of operation is $w = 8$.

## 7   Related Work and Conclusions

In this paper, we formally defined the minimum $w$-cutset problem applicable to any reasoning problems with graphical models such as constraint networks and Bayesian networks. The minimum $w$-cutset problem extends the minimum cycle-cutset problem corresponding to $w = 1$. The motivation for finding a minimal $w$-cutset is to bound the space complexity of the problem (exponential in the width of the graph) while minimizing the required additional processing time (exponential in the width of the graph plus the size of cutset).

The cycle-cutset problem corresponds to the well-known weighted vertex-feedback set problem and can be approximated within factor 2 of optimal by a polynomial algorithm. We show that the minimal $w$-cutset problem is harder. Since any set multi-cover problem (SMC) ([20]) can be reduced to a $w$-cutset problem of a clique tree, the $w$-cutset problem of a tree-decomposition is at least as hard as SMC and cannot have a constant-factor polynomial approximation algorithm unless P=NP. This complexity result applies to the problem of finding minimal $w$-cutset of a graph. Consider chordal graphs discussed earlier in Section 2.1. Each chordal graph has a corresponding tree decomposition whose clusters are the maximal cliques of the graph and vice-versa. If there was a constant-factor $w$-cutset algorithm for a graph, then it could be used to find the minimal $w$-cutset of a chordal graph that is the minimal $w$-cutset of the corresponding tree-decomposition.

Empirically, we show that the minimal cycle-cutset heuristics based on the degree of a node is not competitive with the tree-decomposition of the graph. We also demonstrate that the proposed GWC-family algorithms consistently outperform the previously defined algorithms based on the node elimination order in [18, 13] and [11]. In [18, 13], the next elimination node is added to the cutset if its bucket size exceeds the limit. A similar approach was explored in [11] in DGR algorithm (presented in the empirical section) except that the cutset node was chosen heuristically among all the nodes that were not eliminated yet. The immediate drawback of either approach is that it does not permit to change the order of the nodes already eliminated. As the empirical results demonstrate, DGR usually finds smaller cutset than MGA but bigger than GWC/GWCA/GWCD.

The main objective of our future work is to find good heuristics for $w$-cutset problem that are independent from the tree-decomposition of a graph. So far, we only looked at the degree of a node as a possible heuristics and found empirically that GWC heuristics are usually superior. Alternatively, we want to incorporate into the heuristic measure of the "goodness" of the $w$-cutset other factors such as correlation between variables that strongly affect the convergence of the Markov-chain based $w$-cutset sampling. Finally, there are open questions remaining regarding the relationship between $w$-cutset of a graph and a $w$-cutset of its tree-decomposition. It is not clear, for example, whether the minimal $w$-cutset of a graph is a $w$-cutset of one of its minimum width tree-decompositions.

## 8 Acknowledgements

This work was supported in part by NSF grant IIS-0086529 and MURI ONR award N00014-00-1-0617.

## References

[1] S. A. Arnborg, 'Efficient algorithms for combinatorial problems on graphs with bounded decomposability - a survey', *BIT*, **25**, 2–23, (1985).

[2] A. Becker, R. Bar-Yehuda, and D. Geiger, 'Random algorithms for the loop cutset problem', in *Uncertainty in AI*, (1999).

[3] A. Becker and D. Geiger, 'A sufficiently fast algorithm for finding close to optimal junction trees', in *Uncertainty in AI*, pp. 81–89, (1996).

[4] M. Bellare, C. Helvig, G. Robins, and A. Zelikovsky, 'Provably good routing tree construction with multi-port terminals', in *Twenty-Fifth Annual ACM Symposium on Theory of Computing*, pp. 294–304, (1993).

[5] B. Bidyuk and R. Dechter, 'Cycle-cutset sampling for bayesian networks', *Sixteenth Canadian Conf. on AI*, (2003).

[6] B. Bidyuk and R. Dechter, 'Empirical study of w-cutset sampling for bayesian networks', *UAI*, (2003).

[7] R. Dechter, 'Enhancement schemes for constraint processing: Backjumping, learning and cutset decomposition', *Artificial Intelligence*, **41**, 273–312, (1990).

[8] R. Dechter, 'Bucket elimination: A unifying framework for reasoning', *Artificial Intelligence*, **113**, 41–85, (1999).

[9] R. Dechter, *Constraint Processing*, Morgan Caufmann, 2001.

[10] R. Dechter and J. Pearl, 'Network-based heuristics for constraint satisfaction problems', *Artificial Intelligence*, **34**, 1–38, (1987).

[11] D. Geigher and M. Fishelson, 'Optimizing exact genetic linkage computations', in *7th Annual International Conf. on Computational Molecular Biology*, pp. 114–121, (2003).

[12] U. Kjaerulff. Triangulation of graphs - algorithms giving small total space, 1990.

[13] J Larossa and R. Dechter, 'Dynamic combination of search and variable-elimination in csp and max-csp', *Constraints*, (2003).

[14] S.L. Lauritzen and D.J. Spiegelhalter, 'Local computation with probabilities on graphical structures and their application to expert systems', *Journal of the Royal Statistical Society, Series B*, **50(2)**, 157–224, (1988).

[15] C. Lund and M. Yannakakis, 'On the hardness of approximating minimization problems', *J. of ACM*, **41**(5), 960–981, (September 1994).

[16] J. Pearl, *Probabilistic Reasoning in Intelligent Systems*, Morgan Kaufmann, 1988.

[17] S. Rajagopalan and V.V. Vazirani, 'Primal-dual rnc approximation algorithms for (multi)set (multi)cover and covering integer programs', *SIAM J. of Computing*, **28**(2), 525–540, (1998).

[18] I. Rish and R. Dechter, 'Resolution vs. search; two strategies for sat', *J. of Automated Reasoning*, **24(1/2)**, 225–275, (2000).

[19] Luca Trevisan, 'Non-approximability resu.lts for optimization problems on bounded degree instances', *In proceedings of 33rd ACM STOC*, (2001).

[20] V. V. Vazirani, *Approximation Algorithms*, Springer, 2001.